# General Rules for the Impact of Energetic Disorder and Mobility on Nongeminate Recombination in Phase-Separated Organic Solar Cells

*Guangzheng Zuo[1], Safa Shoaee[1], Martijn Kemerink[2]\*, Dieter Neher[1]\**

[1]Dr. G. Zuo, Prof. S. Shoaee, Prof. D. Neher, Institute for Physics and Astronomy, University of Potsdam, 14476 Potsdam-Golm, Germany
E-mail: martijn.kemerink@cam.uni-heidelberg.de, neher@uni-potsdam.de

[2]Prof. Dr. M. Kemerink
Centre for Advanced Materials, Ruprecht-Karls-Universität Heidelberg, 69120 Heidelberg, Germany

**ABSTRACT**

State of the art organic solar cells exhibit power conversion efficiencies of 18 % and above. These devices benefit from the suppression of free charge recombination with regard to the Langevin-limit of charge encounter in a homogeneous medium. It has been recognized that the main cause of suppressed free charge recombination is the reformation and resplitting of charge transfer states at the interface between donor and acceptor domains. Here, we use kinetic Monte Carlo simulations to understand the interplay between free charge motion and recombination in an energetically-disordered phase-separated donor-acceptor blend. We identify conditions for encounter-dominated and resplitting-dominated recombination. In the former regime, recombination is proportional to mobility for all parameters tested and only slightly reduced with respect to the Langevin limit. In contrast, mobility is not the decisive parameter determining the non-geminate recombination coefficient $k_2$ in the latter case where $k_2$ is a sole function of the morphology, CT and CS energetics and CT states decay properties. Our simulations also show that free charge encounter in the phase-separated disordered blend is determined by the average mobility of all carriers, while CT reformation and resplitting involves mostly states near the transport energy. Therefore, charge encounter is more affected by increased disorder than the resplitting of the CT state. As a consequence, for a given mobility, larger energetic disorder in combination with a higher hopping rate is preferred. These findings have important implications for the understanding of suppressed recombination in solar cells with non-fullerene acceptors which are known to exhibit lower energetic disorder than fullerenes.

**Introduction**

The power conversion efficiency of organic solar cells (OSC) has considerably improved over the past years, now surpassing 18%. [1–5] Due to the introduction of highly absorbing non-fullerene acceptors (NFAs), the short circuit current density, $J_{SC}$, of such cells now approaches the values delivered by more traditional semiconductors of comparable bandgap. Still, OSCs lag behind in terms of the open circuit voltage, $V_{OC}$, and the fill factor, $FF$, the latter being determined by the competition between free charge extraction and recombination. It is well documented that free charge recombination, also denoted as non-geminate recombination, NGR, in organic solar cells is of second order nature: [6]

$$R = k_2 n_e n_h = k_2 n_{CS}^2 \qquad (1)$$

Here, $R$ is the volume recombination rate, $k_2$ the second order NGR coefficient, and $n_e$ and $n_h$ is the density of free electrons and holes, respectively. For intrinsic semiconductors $n_e = n_h \equiv n_{CS}$, the density of charge separated states. For bimolecular free charge recombination, the competition between charge extraction and recombination can be described by analytical "figure of merits" (FOMs) [7], one of these being:

$$\alpha^2 = \frac{q^2 k_2 G d^4}{4\mu_e \mu_h (k_B T)^2}. \qquad (2)$$

Here, $q$ is the elementary charge, $G$ the volume photogeneration rate, $d$ the active layer thickness. $\mu_e$ and $\mu_h$ is the electron and hole mobility, respectively, and $k_B T$ the thermal energy. It was shown that for $\alpha < 1$ the *JV*-characteristics approach the traditional Shockley-diode equation while the photocurrent becomes transport-limited for $\alpha > 1$. [7] Therefore, strategies to reduce $k_2/\mu_e \mu_h$ are desirable and form the motivation to this work.

In the simplest case of encounter-dominated recombination in a homogenous medium, NGR is described by the Langevin recombination coefficient ($k_L$)

$$k_2 = k_L \equiv q(\mu_e + \mu_h)/\varepsilon_0 \varepsilon_r \tag{3}$$

rendering $k_2$ proportional to the sum of the electron and hole mobilities. Here, $\varepsilon_0$ and $\varepsilon_r$ is the vacuum and relative dielectric constant, respectively. The $k_2$ of bulk heterojunction (BHJ) OSCs is generally much smaller, corresponding to a more than 100 times suppression of recombination compared to the Langevin limit in some extreme cases. [8–15] Such BHJ donor-acceptor blends ideally consist of extended domains of the neat donor and acceptor components. It was proposed that stabilization of free charges on these domains reduces the rate through which oppositely charged free carriers meet at the donor-acceptor-interface [15–17]. As outlined below, kinetic Monte Carlo (kMC) simulations showed that phase separation is not sufficient to explain the highly suppressed recombination found in some BHJ OSCs [17,18]. In 2015, Burke et al proposed that NGR is the consequence of efficient resplitting of CT states into free charges [19]. In this picture, free charge recombination results in the formation of interfacial CT states, which either decay back to the ground state or re-dissociate into free carriers (Figure 1). Because the CT dissociation rate itself is a function of carrier mobility, the direct proportionality between $k_2$ and $\mu$ as in the Langevin limit may not hold true anymore. Another feature of the active layer of OSCs is the intrinsic energetic disorder. Energetic disorder is known to slow down carrier motion and to lead to transient (dispersive) phenomena. Various theoretical experimental work studied the effect of disorder on characteristic photovoltaic parameters or on the dispersive recombination of free charges [20–28]. In most of these studies, the focus was on the interrelation between characteristic parameters and the shape of the DOS (exponential or Gaussian) or the predominant recombination pathway. Fewer studies dealt with the interplay between free charge motion and recombination. An elegant analytical approach to relate these two processes was developed by Orenstein and Kastner [20,29]. These authors assumed that charges move primarily at states at a transport energy $E_{tr}$ and that charges situated in the tail of the inhomogeneously broadened DOS must be excited back to that transport energy to be able to meet the countercharge. Since trapping and detrapping also governs the drift of carriers in an electric field, recombination and mobility are expected to follow similar dependencies on time and temperature and carrier density [30]. However, these and related theoretical approaches are applicable only to homogeneous media. Grove [17] and later Heiber [18] applied kMC to relate carrier transport in a phase-separated morphology to encounter-limited recombination. For the simplest case of balanced mobilities, encounter-limited NGR recombination rate was shown to be proportional to $k_L$, and only slightly reduced (by no more than 50 %) compared to the Langevin limit. Interestingly, energetic disorder had an only weak effect on the suppression of encounter-limited recombination in a phase-separated morphology despite its large influence on charge carrier mobility. Very recently, Coropceanu et al. studied the transient NGR in phase-separated donor-acceptor blends under explicit consideration of CT formation, resplitting and decay [31] . This study showed a significant attenuation of $R$ with increasing disorder and CT lifetime, but no correlation to the carrier mobility was made.

Here, we perform kMC simulations of the free carrier recombination in disordered phase-separated blends. By varying the attempt-to-hop frequency $v_0$ and the energetic disorder $\sigma$ over a wide range, we evaluate in detail the interplay between free charge carrier motion and recombination. Doing these simulations in a transient fashion we gain access to the extraction-recombination balance as a function of the mean energy of the free carriers, down to the equilibrium situation. For low $v_0$, recombination is encounter-dominated and proportional to the Langevin recombination rate, though reduced by a geometric reduction factor which is nearly independent of energetic disorder. Increasing $\nu_0$ causes a transition to the resplitting-dominated regime, where CT exciton dissociation competes efficiently with its decay to the ground state. Importantly, in this regime the NGR coefficient becomes independent of $\nu_0$, and for a given energetic disorder, independent of $\mu$. Here, $k_2$ is given only by the details of the energetics and morphology of the blend. Our simulations also show that the formation of a CT exciton in a phase-separated blend involves the thermal excitation of both the electron and the hole to their respective transport energies. As a consequence, charge encounter is more affected by increased disorder than the resplitting of the CT state, causing a recombination reduction factor as small as $10^{-3}$ for high disorder and high attempt-to-hop frequencies. Such conditions were reported for fullerene-based blends; being a possible reason for the highly suppressed recombination in some fullerene-containing bulk heterojunction solar cells.

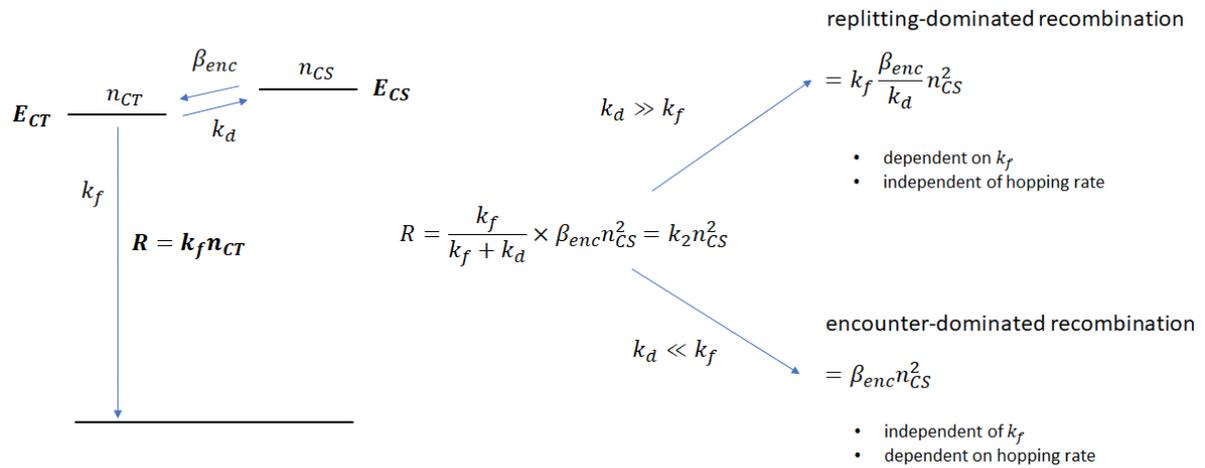

*Figure 1*: *Scheme showing the states involved in free charge recombination. Here, $n_{CS}$ and $n_{CT}$ is the density of free carriers (charge separated states) and of charge transfer states, respectively, $\beta_{enc}$ the coefficient for bimolecular free charge encounter forming the CT state, $k_d$ the rate coefficient for CT dissociation into free charges and $k_f$ the rate coefficient for the decay of the CT state to the ground state. Also depicted are the two extremes. For $k_d \ll k_f$,*

*resplitting of the CT state is very unlikely and the recombination rate R is determined by the encounter rate which is independent of the CT decay rate but proportional to the carrier hopping rate and with this to the carrier mobility. In contrast, if $k_d \gg k_f$, CT states resplit at a much higher rate than they decay to the ground state. Because $\beta_{enc}$ and $k_d$ depend linearly on the hopping rate $\nu_0$, it's effect on the recombination rate cancels out while on the other hand, R becomes strictly proportional to $k_f$.*

**Theory**

In the following, we briefly review free charge recombination through the formation and decay of CT states. The states and rates involved in this process are depicted in Scheme 1. A more detailed treatment including the reformation of singlet states or the build-up of triplet population has been treated in detail in [33] and [34], but the inclusion of these states does not change the general conclusions.

In absence of generation, the kinetics of the excited state population is determined by the following two differential equations [19]:

$$\frac{dn_{CS}}{dt} = -\beta_{enc} n_{CS}^2 + k_d n_{CT}, \tag{4a}$$

$$\frac{dn_{CT}}{dt} = \beta_{enc} n_{CS}^2 - (k_d + k_f) n_{CT}. \tag{4b}$$

We now consider the case of a dynamic equilibrium between the free charges and CT states, where $\frac{dn_{CT}}{dt}$ is smaller than the rates of the processes that increase or decrease the CT state population, as expressed by $\beta_{enc} n_{CS}^2 \cong (k_d + k_f) n_{CT}$. In this limit, R can be related to the free carrier density $n_{CS}$ via:

$$R = k_f n_{CT} = \frac{k_f}{k_d + k_f} \beta_{enc} n_{CS}^2 = \gamma_{CT} \beta_{enc} n_{CS}^2 \equiv k_2 n_{CS}^2. \tag{5}$$

Here,

$$\gamma_{\text{CT}} = \frac{k_f}{k_d+k_f}. \tag{6}$$

is the recombination reduction factor due to resplitting of the CT state.

For $k_d \ll k_f$, $\gamma_{\text{CT}} \cong 1$ and $R = R_{enc} = \beta_{enc}n_{\text{CS}}^2$, meaning that every charge pair forming a CT state will recombine. We denote this as the *encounter-dominated* recombination regime. It is common to relate $\beta_{\text{enc}}$ to the charge carrier mobility via

$$\beta_{enc} = \gamma_{\text{enc}}k_L, \tag{7}$$

where $\gamma_{\text{enc}}$ ($\leq 1$) is the reduction factor due to the confinement of opposite charges in their nano-domains in a BHJ system (also called the geometric reduction factor). As a consequence, $k_2$ and with this $R$ depend linearly on the sum of the charge carrier mobilities. For disordered systems, the mobility is given by the product of the carrier hopping frequency $\nu_0$ and a term describing the distribution of the carriers in space and energy. For a regular lattice and fully equilibrated carriers, this term becomes proportional to $exp\left\{-\frac{4}{9}\left(\frac{\sigma_{DOS}}{k_BT}\right)^2\right\}$ where $\sigma_{DOS}$ is the width of the Gaussian disorder and $k_BT$ the thermal energy. [35,36] Therefore, in the encounter-dominated case, $k_2$ will be linear in $\nu_o$ and decrease rapidly with increasing $\sigma$, while not depending on the CT decay rate $k_f$.

In the other limit, $k_d \gg k_f$ and ($\gamma_{CT} \ll 1$). We denote this as the *resplitting-dominated* recombination regime. In this case, $\gamma_{\text{CT}} \cong k_f/k_d$ and Eq. 5 becomes:

$$k_2 = k_f \frac{\beta_{\text{enc}}}{k_d} = k_f \frac{\gamma_{enc}k_L}{k_d}. \tag{8}$$

Here, the most important parameter determining the suppression of recombination is $\frac{\gamma_{enc}k_L}{k_d}$, which expresses the interplay between CT formation and resplitting. In absence of energetic disorder, results of kMC simulations on a lattice with lattice parameter $a$ can be approximated by [37]

$$k_d = Bk_L \frac{3}{4\pi a^3} exp\left(-\frac{E_{CS}-E_{CT}}{k_B T}\right) \quad (9)$$

where $B$ is a prefactor depending on the lattice and interface morphology. For charge separation in a homogeneous medium, $B \cong 0.5$, while $B = 5\sim15$ was found to give the best fit to the efficiency of charge separation a phase-separated morphology as simulated by kMC (note that in the sentinel paper by Braun, $B = 1$). Irrespective of these details, Eq. 9 in combination with Eq. 8 predicts that $k_2$ becomes independent of the carrier mobilities for efficient CT resplitting due to the cancelation of $k_L$, but that it is largely determined by the energy and site densities of the contributing states. Clearly, the situation becomes more difficult in presence of disorder where there is a distribution of energies for both CT states and free charges. This is the main topic of this work.

**Kinetic Monte Carlo modelling**

The kMC algorithm has been introduced in previous works. [38–43] Details can be found in the Supporting Information. In brief, we consider a cubic lattice with lattice parameter $a_{NN} = 1.8\ nm$, on which charges can hop to a fixed number of neighboring sites. For the current simulations we used 26 neighbors, which corresponds to a 3x3x3 cube around each site. Hopping rates are calculated from the Miller-Abrahams expression with an attempt-to-hop frequency $\nu_0$ that has the physical meaning of the success rate of downward hops to a nearest neighbor site. For upward hopping, the rate is reduced accordingly. Following previous literature [44–46], we considered the range $\nu_0 = 10^8 - 10^{12}\ s^{-1}$, which is large enough to clearly identify the different recombination regimes that are the topic of this work.

A statistical Gaussian disorder of width $\sigma_{DOS} = 65, 90$ or $120\ meV$. These values were chosen because they correspond to a stepwise 10 times reduction of the steady state mobility at $T = 300\ K$ while representing the range of disorder values reported for carrier transport in BHJ OSCs. For charge transport and recombination in fully homogeneous media, an effective hopping medium is defined with electron and hole hopping parameters and energies that correspond to the donor HOMO and acceptor LUMO level, respectively. Hereby, each site can function as a donor and acceptor. An on-site barrier for electron-hole encounter is implemented to introduce a driving force for charge transfer and to avoid direct exciton formation. For phase-separated systems, the morphology is implemented by assigning individual hopping sites to different material phases. Here, we used a simplified phase separated morphology consisting of 7x7 columnar inclusions in a 10x10 columnar unit cell, where the column axis is in the direction of charge extraction. Electron transfer to the acceptor phase is driven by a LUMO level offset $\Delta E^{LUMO} = E_{donor}^{LUMO} - E_{acceptor}^{LUMO} = 600\ meV$ (and for holes by a respective HOMO offset of $400\ meV$) to ensure the absolute separation of hole-electron pairs on site, as well as referring a common LUMO energies of -3.5 eV, -4.1 eV and HOMO of -5.5 eV, -5.9 eV for donor and acceptor, respectively. In recent work, this morphology was demonstrated to be sufficient to describe device current-voltage characteristics. [46] We consider random energetic disorder and the same width of the DOS at the interface and in the bulk. We acknowledge that this approximation oversimplifies the energetics in a real BHJ blend with crystallized donor and acceptor domains and an intermixed interface. The simulation of such a complex morphology is beyond the scope of our paper, which is to establish fundamental correlations between energetic disorder, carrier mobility and NGR. With the same argument, we refrain from introducing charge or CT delocalization.

**Results**

Figure 2a shows an example of the temporal decay of $n_{CT}$, $n_{CS}$ and the number of total charges (electron-hole pairs) $n = n_{CT} + n_{CS}$. The parameters have been chosen to reproduce carrier densities and lifetimes as typically reported for OSCs under steady state illumination conditions [27,47–49]. Because we start with free charges distributed homogeneously on all donor and acceptor sites, the first step is the formation of an appreciable CT population due to free charge encounter, which for the phase-separated morphology requires the charges to meet at the DA heterojunction. Concurrently, the free carrier density decreases while the total number of charges remains rather constant during this process. This initial regime is followed by the decay of the carrier population due to the decay of populated CT states to the ground state. Importantly, in this regime, $\frac{dn_{\text{CT}}}{dt} \ll \frac{dn_{\text{CS}}}{dt}$, meaning that dynamic equilibrium determines the CT state population and the above considerations can be applied. As expected and shown before, increasing the hopping rate $\nu_0$ speeds up recombination as charge encounter becomes more frequent, as seen by the continuous increase of the free charge NGR coefficient $k_2$ with $\nu_0$ in Figure 2b. However, this increase becomes less pronounced at high $\nu_0$, indicating a transition to *resplitting-dominated* recombination.

The direct comparison of the carrier and CT kinetics for different $\nu_0$ is, however, difficult because the larger is $\nu_0$, the faster is the equilibration of photogenerated charge carriers in their respective DOS. A way out is to plot the data as function of the reduced time $t \times \nu_0$, as documented for the time dependence of the corresponding mobility in Figure S1. Notably, when we additionally divide the mobility by $\nu_0$, the traces for different $\nu_0$ fall onto one line, despite the different recombination properties for low and high hopping rates. This indicates that recombination does not concern predominately the faster carriers but rather affects the entire carrier population. Figure 3 compares $\mu(t \times \nu_0)$ to $k_2(t \times \nu_0)$ for $k_f = 10^8\ s^{-1}$ (a) and $k_f = 10^7\ s^{-1}$ (b). We are aware of the fact that a CT decay rate of $10^7\ s^{-1}$ is at the very upper end of the reported values. As pointed out earlier, we have deliberately chosen a simple lattice model without charge or CT delocalization. This causes a fairly high CT binding energy $E_{bind} = E_{CS} - E_{CT} = 228\ meV$ at $T = 300\ K$ for the chosen parameters and, concurrently, a small dissociation rate. However, all the results and conclusions from this paper can be easily applied to the experimental situation by proper scaling of all rates. Note that in Figure 3, the transient mobility data have been aligned to the transient $k_2$ according to the encounter-limited case, where $k_2 = \gamma_{enc} k_L = \gamma_{enc} 2q\mu/\varepsilon_0\varepsilon_r$, thereby taking into account a slight suppression of encounter in the phase-separated morphology with $\gamma_{enc} = 0.4$.

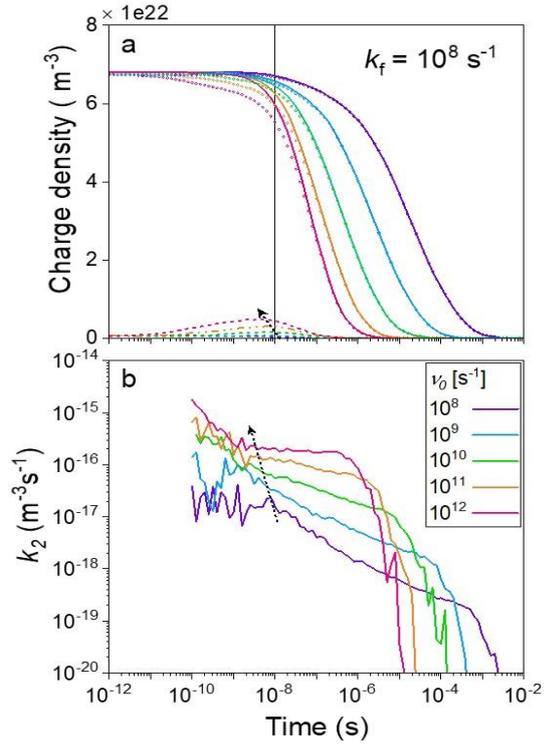
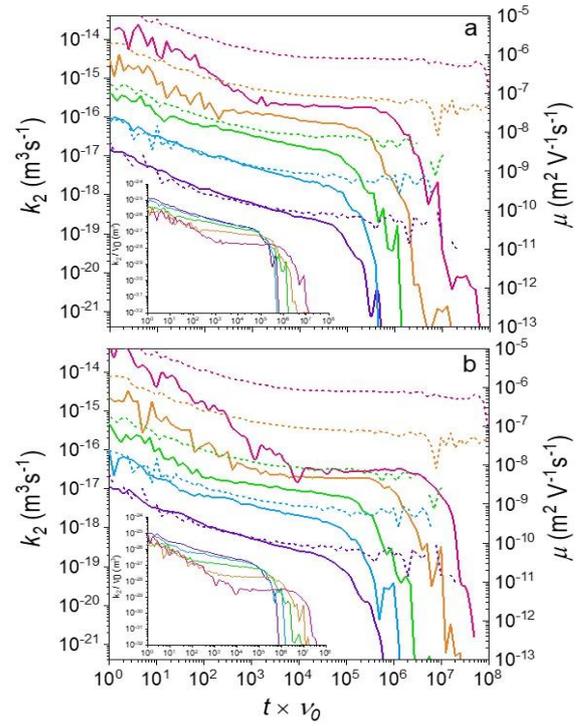

*Figure 2*: kMC results of the dynamics of the total density of electron-hole pairs (solid lines), of free electron-hole pairs (dots) and of CT states (dashed lines) for variable attempt-to-hop frequencies ($v_0$) *(a)* and the corresponding second order NGR coefficient ($k_2$) *(b)*. The dotted arrow indicates the maximum of the CT population. Simulations were performed with the following parameters: $\sigma_{DOS,e} = \sigma_{DOS,h} = 90$ meV, $k_{CT}$ ($k_f$) = 1×10$^8$ s$^{-1}$, T = 300K, intersite distance ($a_{NN}$) = 1.8nm, $v_{0,e} = v_{0,h}$ = 1×10$^8$, 10$^9$, 10$^{10}$, 10$^{11}$ and 10$^{12}$ s$^{-1}$ (purple to red).

*Figure 3*: Bimolecular recombination coefficient (solid lines) and charge carrier mobility (dashed lines) for variable attempt-to-hop frequencies at $k_f$ = 1×10$^8$ s$^{-1}$ *(a)* and 1×10$^7$ s$^{-1}$ *(b)*, plotted as function of $t \times v_o$. In both graphs, the two y-axes have been aligned according to the encounter-dominated case with $\gamma_{enc} = 0.4$. The insert panels show the corresponding $k_2/v_0$ transients. kMC parameters are the same as in *Figure 2*.

For low $v_0$, $k_2$ is indeed dominated by charge encounter and independent of $k_f$. One the other hand, the system enters the re*splitting-dominated* regime for high enough $v_0$, where now $k_2$ becomes proportional to $k_f$ but independent of $\nu_0$ (and $\mu$). In the most extreme case treated in Figure 3, recombination is hundred times suppressed. The transition from *encounter-* to *resplitting-dominated* recombination is more clearly seen in the insets where we plot $k_2/\nu_0$ as function of $t \times \nu_0$, and where all curves with *encounter-dominated* recombination fall onto one line.

Figure 4 provides a more detailed look at the carrier dynamics for the case of $\sigma_{DOS} = 90\ meV, k_f = 10^8\ s^{-1}$ (see Figure S2 for the same data but plotted as function of time). As expected from the earlier observation that $\mu/\nu_0\ (t \times \nu_0)$ is independent of $\nu_0$, we observe very similar values and transients of the mean energy of the charges (a) and of the energy of moving charges (b) for the different hopping rates when plotted as function of $t \times \nu_0$. As extensively discussed in the literature, the mean energy of the carriers continuously decreases and eventually approaches the energy of the equilibrated carriers, $E_\infty$, which for electrons (holes) is $\sigma_{DOS}^2/k_B T$ below (above) the centre of the respective DOS. Because we start with free carriers distributed among the whole cell volume, the mean energy of the electrons (holes) is initially above (below) the centre of the acceptor LUMO (donor HOMO). The blue vertical line marks the time when the mean energy crosses the effective transport energy ($E_{tr} = E_{LUMO} - \sigma_{DOS}^2/2k_B T$ for electrons and $E_{HOMO} + \sigma_{DOS}^2/2k_B T$ for holes). Once the carriers sink deeper into the DOS, their motion requires thermal excitation back to $E_{tr}$, as demonstrated by the saturation of the energy of the moving carriers in panel b. Consequently, the carrier mobility reduces continuously even beyond this time (panel c). In contrast, the mean energy of the recombining CT states shows a sudden transition to a nearly constant value once the carrier package crosses $E_{tr}$ (see panel d). Interestingly, this plateau is around an energy of $E = E_{G,tr} - E_{bind} = \left(E_{LUMO} - \frac{\sigma_{DOS,e}^2}{2k_B T}\right) - \left(E_{HOMO} + \frac{\sigma_{DOS,h}^2}{2k_B T}\right) - E_{bind}$, which is the effective transport gap, $E_{G,tr}$, minus the CT binding energy. This suggests that once the carriers have moved deep into their respective DOS, CT states form predominately by the encounter of electrons and holes that are both traveling at their respective $E_{tr}$, while there is little contribution by equilibrated charges at $E_\infty$. A similar situation is encountered at lower and higher disorder values (Figure S3 and S4), where a transition to a nearly constant CT energy is seen when the carrier package crosses $E_{tr}$, though the transition is more smeared out for high disorder. In all cases, the data assemble around $E = E_{G,tr} - E_{bind}$ at long enough times after photogeneration (see Figure S5 for the direct comparison of the three cases). It has been suggested that CT formation in organic blends requires thermal excitation of at least one of the meeting carriers up to the transport level [39,50] Our data show that in phase-separated blends, thermal excitation of both encountering carriers is involved in CT formation [30]. The situation is different in a fully intermixed system with otherwise identical parameters (Figure S6). Here, we observe a pronounced time dependence of the CT emission energy for all hopping rates, moving well below $E_{G,tr} - E_{bind}$ throughout the entire recombination process. This indicates that recombination between free and trapped carriers becomes the dominant channel for CT formation at longer times. Another important feature is the slight down-shift of the CT energy with increasing $\nu_0$. This shift occurs independent of whether recombination is encounter-dominated (every formed CT state recombines) or suppressed through CT-resplitting. We propose that this process is mainly due to CT energy relaxation, though we cannot rule out contributions from the preferential resplitting of higher energy CT states for large $\nu_o$. In both cases, the extent of the red-shift is expected to increase with $\nu_o/k_f$. This is well documented by the comparison of the CT transients for $k_f = 10^8 \ s^{-1}$ and $k_f = 10^7 \ s^{-1}$ in Figure S6, where we observe the very same final energies for the same $\nu_o/k_f$.

Figure 4e displays the time dependence of $k_2$ while Figure 4f shows the recombination reduction factor $\gamma = \gamma_{geo} \times \gamma_{CT}$. As discussed earlier, the initial free carrier loss is due to CT formation. Beyond this, the course of $k_2$ follows roughly the time dependence of $\mu$, suggesting $k_2 \propto \mu$ (see also Figure 3). As a result, $\gamma$ is nearly independent of time. While this is expected in the encounter-dominated case, where $\gamma = \gamma_{enc}$, our results suggest that also $\gamma_{CT}$ exhibits an only weak (if any) time dependence, even when $\mu$ is time-dependent. The effect is most pronounced for a high disorder, where a continuous decrease of $k_2$ over time (Figure S4e) contrasts a nearly time-independent reduction factor (Figure S4f) even in the case of strongly suppressed recombination.

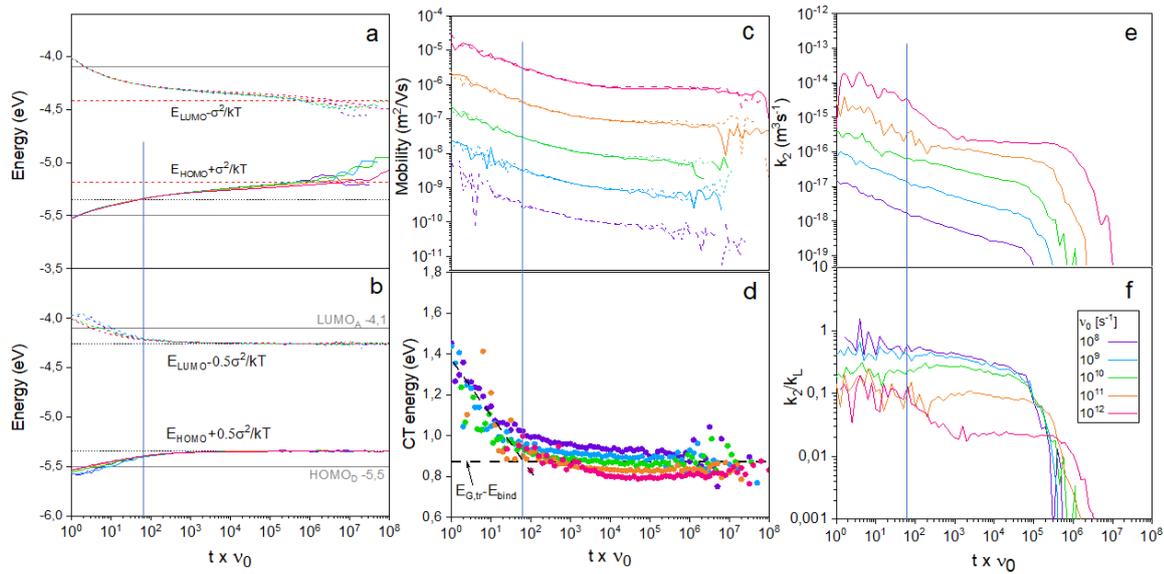

*Figure 4:* *Simulated transients of **(a)** the mean energy of all charges, **(b)** the mean energy of moving charges, **(c)** the bipolar mobilities for holes (solid lines) and electrons (dashed lines), **(d)** the energy of recombining CT states, **(e)** $k_2$ and **(f)** of the recombination reduction factor for variable attempt-to-hop frequencies plotted as function of $t \times v_o$. The kMC parameters are the same as in Figure 2. The blue vertical guideline marks the time when the mean energy crosses the transport energy $E_{tr} = E \mp \sigma_{DOS}^2/2k_BT$ (grey horizontal dashed guideline), and the red horizontal guideline shows the equilibrium energy $E_\infty = E \mp \sigma_{DOS}^2/k_BT$, where $(-)$ for electrons and $(+)$ for holes.*

**Discussion**

Before proposing a model to conclusively describe the simulation results, let us summarize the main findings. First, the data shows that a dynamic equilibrium exists between the mobile charges at $E_{tr}$ and those situated deeper in the DOS for all conditions considered here, and that this equilibrium is not severely disturbed by the formation and decay of the CT states. This is in accordance with the concept of the demarcation energy $E_D$ as introduced by [20,29] where free charges are in quasi-equilibrium with the main carrier distribution situated at $E_D$. As such, the rate at which charges encounter to form the CT state is determined by the mobility averaged over all carriers in the DOS. In accordance with this conclusion is the very weak effect of disorder and CT decay rate on the reduction factor in the encounter-dominated regime (Figure S7). Groves et al. reported a weak decrease of $\gamma_{enc}$ by only 60 % when increasing $\sigma$ from 0 to 75 meV [17]. The observed range of values for $\gamma_{enc}$ of 0.3~0.5 is fully consistent with the results of kMC with a disorder of 75 meV for a comparable domain size as used here [18]. Second, the data show that once the carrier package has relaxed below $E_{tr}$, the mean energy of the recombing CT states is nearly time-independent and close to the effective transport gap minus the CT binding energy. This is suggestive of a situation where CT formation is mainly by charges near $E_{tr}$. In our strict phase-separated blend, CT states are exclusively formed at the sharp DA interface. Motion of the encountering charges in their respective phases towards this interface involves frequent trapping and re-excitation of these charges to $E_{tr}$, where it is more likely that charges meet at the interface when they are in a mobile state. In addition, the an already existing electron (hole) at the DA interface will tilt the DOS of the approaching counter charge, due to the mutual electrostatic interaction. As a consequence, hops to sites higher in the DOS will become more likely [35] – an effect known as carrier heating by a strong electric field. In turn, the initial step of CT dissociation will mainly involve transport sites around the transport energy, leading to a very slow variation of $k_d$ (and with that of $\gamma_{CT}$) with time – if any. This situation is depicted in Figure 5a.

Figure S7 plots the time dependence of $\gamma$ for all cases studied here. Without phase-separation, $\gamma$ is equal to $\gamma_{enc} \cong 1$ in the encounter-dominated regime but it becomes strongly time dependent once recombination becomes suppressed due to CT resplitting. This is explained by the continuous decrease of the CT energy as documented in Figure S6, suggesting also $k_d$ to be time-dependent. In contrast, for all phase-separated systems, there exists a plateau extending over approximately three orders in time (grey dashed line), even for the highly suppressed cases. For large disorder, this plateau is overlaid by a global decrease of $\gamma$ with time, which is however far weaker than the decrease in $\mu$ in the considered time range.

Figure 5b-d plots the reduction factor from these plateau regions versus the hopping rate. Regimes of encounter-dominated and resplitting-dominated recombination are clearly discernible. In the latter regime, increasing the disorder for a given hopping rate increases $\gamma$ (rendering recombination less suppressed) because the CT dissociation slows down and becomes less competitive to the CT dissociation decay. For the same reasons, a slower CT decay (smaller $k_f$) will decrease $\gamma$. In order to get a more quantitative picture, the simulation data were modelled by assuming that $k_d$ is time-independent and (for a given $\sigma_{DOS}$) strictly proportional to $\nu_0$, $k_d = c \times \nu_0$. Here $c$ is a dissociation prefactor which is determined by the details of the blend morphology and energetics but is independent of the kinetic parameters $\nu_0$ and $k_f$. Accordingly,

$$k_2/k_L = \gamma = \gamma_{enc}\gamma_{CT} = \gamma_{enc}\frac{k_f}{k_f+k_d} = \gamma_{enc}\frac{k_f}{k_f+c\nu_0}. \tag{10}$$

with $\gamma_{enc}$ and $c$ being the only fitting parameters. The best fits are shown as solid lines in Figure 5b-d. Despite the simplicity of the model, the course of $\gamma(\nu_0)$ is well captured, with a very small dependence of the dissociation prefactor $c$ on the CT decay rate $k_f$. When $\sigma$ increases from 65 meV to 120 meV, $c$ (and with that $k_d$) decreases by circa one order of magnitude while $\mu$ drops by a factor of 100, meaning that the dissociation rate is less affected by disorder than the global mobility. This is agreement to our model that charge encounter causes an athermal CT population, which is more prone to redissociation than a fully thermalized population.

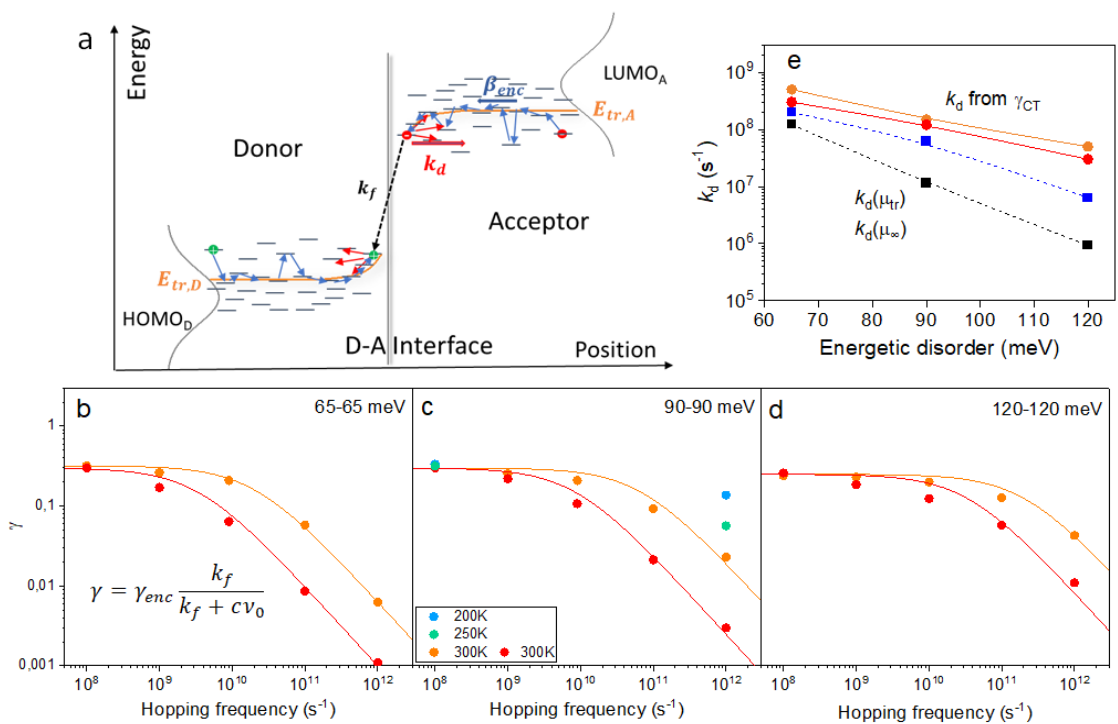

*Figure 5: (a) Scheme of CT formation, resplitting and decay in a phase-separated disordered BHJ system. Charge encounter is determined by a time-dependent encounter rate $\beta_{enc}(t) = \gamma_{enc} k_L(\mu(t))$ involving the entire density of occupied states (DOOS). In contrast, for long enough delay after photoexcitation, the dissociation rate $k_d$ is mainly determined by states around $E_{tr}$. (b, c, d) kMC results of reduction factor γ of (nearly) equilibrated carriers for different hopping rates and disorders in phase-separated blends (taken at the times indicated in Figure S7) for $k_f = 10^8\ s^{-1}$ (orange circle) and $k_f = 10^7\ s^{-1}$ (red circle). Solid lines show the best fit to the data with Eq. 10. All simulations were performed at T=300 K, expect for the*

*green (T=250 K) and blue (T=200 K) dots in Figure (c). (e) $k_d$ as function of energetic disorder as deduced from the fit of the reduction factor in Figure 5b-d for $v_0 = 10^{11} s^{-1}$ and $k_f = 10^8\ s^{-1}$ (see Table 1 for the parameters). Also shown is the analytical prediction of $k_d$ in absence of disorder according to Eq. 9, where µ is either the transient mobility when the carrier package crosses $E_{tr}$ or after the carriers have equilibrated at $E_\infty$ (dotted lines).*

**Table 1**: *$\gamma_{enc}$ and c from the fit of the reduction factor in Figure 5(b,c,d) with Eq. 10 for different energetic disorder and CT decay time.*

| Energetic disorder (meV) | $k_f = 1\times10^8$ s$^{-1}$ | | $k_f = 1\times10^7$ s$^{-1}$ | |
|---|---|---|---|---|
| | c | $\gamma_{enc}$ | c | $\gamma_{enc}$ |
| 65-65 | 5×10$^{-3}$ | 0.32 | 3×10$^{-3}$ | 0.3 |
| 90-90 | 1.5×10$^{-3}$ | 0.3 | 1.2×10$^{-3}$ | 0.3 |
| 120-120 | 5×10$^{-4}$ | 0.25 | 3×10$^{-4}$ | 0.25 |

Figure 5e plots $k_d$ as deduced from the fits in Figure 5 for $v_0 = 10^{11}\ s^{-1}$. Because of the presence of energetic disorder, an analytical model to describe the dependence of $k_d$ on $\sigma_{DOS}$ is unfortunately not available. Instead, we added to Figure 5e the analytical predictions for $k_d$ in absence for disorder (Eq. 9), where for $\mu$ we either inserted the mobility at the time when the carrier package crosses $E_{tr}$ (representative of the energy of carriers dissociating from the CT state) or after the carrier population equilibrated at $E_\infty$ (describing the encounter of equilibrated carriers). Further, $B = 10$, $a = 1.8\ nm$, and $E_{bind} = 228$ meV. We find that $k_d(\mu_\infty)$ displays a much stronger dependence on $\sigma_{DOS}$, lying 2 orders of magnitude below the $k_d$ from the recombination analysis at the largest disorder. In contrast, $k_d(\mu_{tr})$ captures quite well the $\sigma_{DOS}$ −dependence of $k_d$, though it underestimates its value for high disorder. This can be understood by the lowing of the effective dissociation barrier due to energetic disorder, which is well-documented in the literature and is not captured by Eq. 9 [51,52].

Nevertheless, the benefit of increased energetic disorder in providing a larger number of low energy states for free carriers is overcompensated by a reduced mobility, resulting in total in a decrease of $k_d$.

To complement our study, we performed additional simulations at different temperatures with $\sigma_{DOS} = 90\ meV$, $k_f = 10^8\ s^{-1}$ and $\nu_0$ set either to $10^8\ s^{-1}$ (encounter-dominated) or $= 10^{12}\ s^{-1}$ (resplitting-dominated). The resulting transients of $k_L$, $k_2$ and $\gamma$ are shown in Figure S8. Decreasing the temperature from 300 K to 200 K slows down thermalization as seen by more pronounced dispersive effects. Importantly, in the encounter-limited regime, $k_2$ decreases the same way with temperature and time as $\mu$, meaning that $\gamma$ ($= \gamma_{enc}$ in the encounter-dominated case) is barely affected by the temperature (Figure 8a,c). This confirms our interpretation that the rate at which CT states form is governed by the motion of carriers towards the heterojunction and that the final step of CT formation state is not the rate-limiting step. Figure S8 b,d displays the situation for the resplitting-dominated case. Here, the effect of temperature on $k_2$ is much weaker than on $\mu$, causing a ca. 10 times increase of $\gamma$ from 300 K to 200 K. This is the consequence of the need of thermal excitation to dissociate the CT state into free charges (Eq. 9), meaning that CT dissociation competes less efficiently with CT decay at lower temperatures.

Our model has important consequences for the conditions to be met when aiming for strongly suppressed free carrier recombination despite a high carrier mobility. Figure 6 and Figure S9 collect $k_2$, $\mu$ and $\gamma$ for (nearly) equilibrated carriers. For a given disorder, increasing $\nu_0$ (and with that $\mu$) causes a transition from the encounter- to the resplitting-dominated regime (Figure S9). Above this transition, i.e. in the resplitting-dominated regime, a further increase of $\nu_0$ ($\mu$) leaves $k_2$ nearly unaffected. High hopping rates are, therefore, indispensably when aiming at a favourable extraction-recombination balance, in full agreement with the prediction of the simple rate model in section III. Reducing the energetic disorder, e.g. by removing impurities or improving the domain crystallinity, moves this transition to lower $v_0$. In the resplitting-dominated regime, a smaller $\sigma$ increases $k_2$ but to a smaller extent than $\mu$. This is the consequence of the competition between the two processes determining $k_2$, namely the increase of $\beta_{enc}$, which favors CT formation, and a more efficient CT resplitting due to a larger $k_d$. Therefore, for a given value of $v_0$, the extraction-recombination balance benefits from lower disorder. A more complete picture (Figure 6) arises when plotting $k_2$ and $\gamma$ as function of mobility - a combined function of the hopping rate and the energetic disorder -: $\mu_\infty \propto v_0 \, exp\left\{-\frac{4}{9}\left(\frac{\sigma_{DOS}}{k_B T}\right)^2\right\}$. For low mobility, $k_2$ is almost entirely determined by mobility, with little influence of energetic disorder. In contrast, mobility is not a decisive parameter for $k_2$ in the *resplitting-dominated* recombination regime. Instead, $k_2$ becomes a sole function of the morphology, $CT$ and $CS$ energetics and $CT$ states decay properties. Our simulations, therefore, predict a large variation of $k_2$ for a given mobility. This is found experimentally, see the insets of Figure 6. As expected, the scatter of the experimental values is larger than of the kMC data in Figure 6, which is due to the fact that other parameters as considered in our simulations (interface morphology, CT delocalization) contribute to $k_d$ and with that to $k_2$. As a consequence, it should be possible to decrease the FOM $\alpha$ in Eq. 2 towards the Shockley-regime ($\alpha < 1$) by independently reducing $k_2$ and increasing $\mu$ through proper design of the molecular and blend structure.

Noticeably, for a given mobility, our simulations show that the larger is $\sigma_{DOS}$, the smaller is $k_2$ and the stronger will $CT$-resplitting reduce the recombination. This is again a direct consequence of the dependence of $k_2$ on $k_d$ and $\beta_{enc}$, where increasing disorder slows down $\beta_{enc}$ more strongly than $k_d$. In turn, if we consider $k_2$ for a given mobility, a larger disorder and hopping rate is desirable. Soluble fullerenes such as PCBM are known to exhibit considerable energetic disorder while also having large hopping rates. [38,42,43,53,54] For example, the analysis of the transient carrier collection in the blend of PC$_{71}$BM with the donor polymer TQ1 with kMC yielded $\sigma_{DOS,e} = 125\ meV$ and $\nu_e = 1.8 \times 10^{13}\ s^{-1}$. *Substituted fullerenes such as PCBM stand out by fairly small aggregate sizes even in well phase-separated blends* [55]*, while they exhibit significant orientational and conformational disorder even within these crystallites* [56]*. Combined with dynamic disorder due to electron-vibration coupling, values of σ as large as 100 meV have been predicted for crystalline PCBM clusters* [57]*. On the other hand, the close distance and near spherical shape of the fullerene conjugated orbitals result in large electron coupling, explaining the overall high electron mobility* [56,58]*.* Modern NFA as used in state of the art OSCs exhibit highly crystalline layers with energetic disorders of typically 50-70 meV [48,59–61]. Unfortunately, only few studies concerned the interplay between free carrier recombination and motion in NFA-based blends, [48,62–64] and even fewer included the measurement of energetic disorder [48,49,59,61]. These studies yielded a fairly wide spread of mobilities and recombination coefficients even for the same material combination, which may in part be related to the details of the molecular structure (e.g. the polymer molecular weight [64,65] etc.) or the layer morphology (e.g. interface properties [31,66]). Therefore, experimental studies on well-defined morphologies are urgently needed to disentangle the different processes governing the mobility-recombination interplay in high performance NFA-based systems.

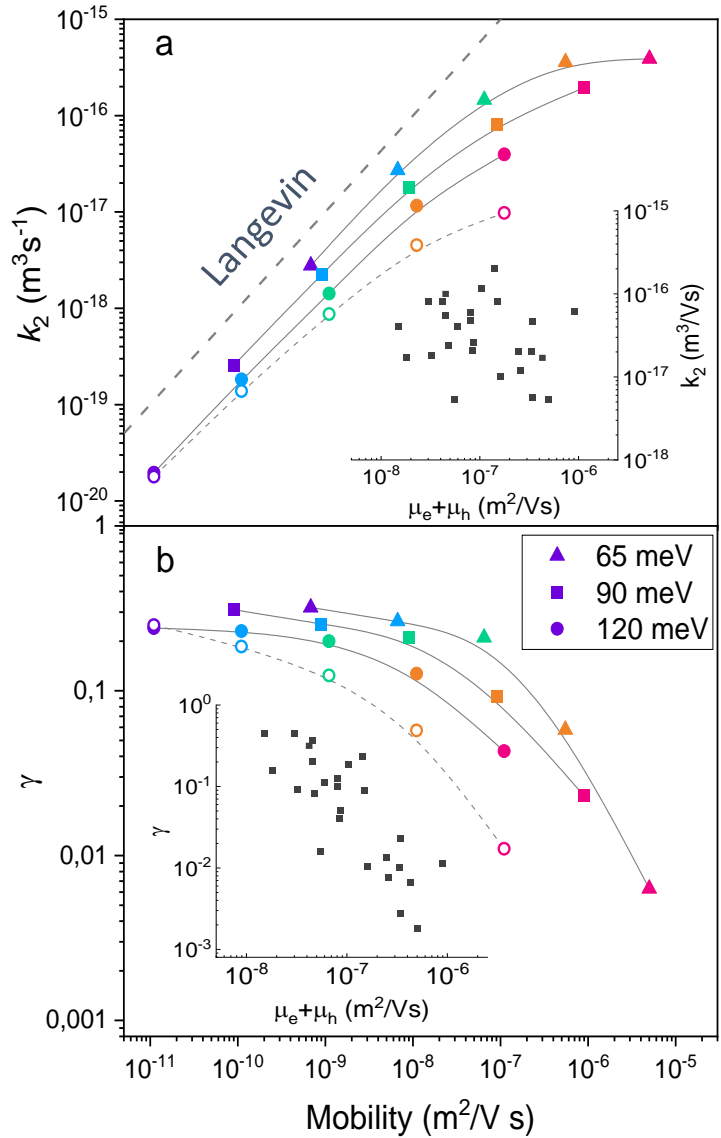

*Figure 6*: Simulated $k_2$ as function of $\mu_\infty$ for $\sigma_{DOS}$ of 65 meV (triangle), 90 meV (square) and 120 meV (circle) for (nearly) equilibrated charges taken at the times indicated in Figure S7). *(a)* and the corresponding deduced reduction factor *(b)*. Solid symbols are for $k_f = 10^8 s^{-1}$ while open circles show the result for $k_f = 10^7 s^{-1}$ for the example of high energetic disorder. The dashed line is the prediction of the Langevin model the solid line guides the eye The inset show experimental $k_2$ and $\gamma$ data as function of $\mu_e + \mu_h$ of the bulk heterojunction solar cells taken from [34] *(grey dots)*.

**Conclusion**

In conclusion, we have employed kinetic Monte Carlo simulations to understand transient free carrier motion and recombination in energetically disordered phase-separated donor-acceptor blends. We find that the bimolecular recombination coefficient can be described by the competition between free charge encounter, CT state resplitting and CT decay, as summarized by $k_2(t) = \frac{k_f}{k_f + c\nu_0} \gamma_{enc} k_L(t)$. Here, $c$ describes the proportionality between the CT dissociation rate and the attempt-to-hop frequency $\nu_0$, and depends only on the energetics and structure of the phase-separated system but not on the details of the free carrier distribution in the DOS. This is a direct consequence of that fact that free charge encounter in the considered phase-separated disordered DA blend forms an athermal excited CT state distribution, whose energy is independent of the mean energy of the free carriers once the carrier population has moved below the effective transport energy. Despite this, the formation of the CT state is strictly proportional to the mobility averaged over all carriers as expressed by the Langevin encounter coefficient $k_L$. For high enough $\nu_0$, the recombination coefficient becomes nearly independent of the attempt-to-hop frequency and, for a given energetic disorder, on the value of the mobility. As such, mobility is no more a decisive parameter for the value of $k_2$ in the resplitting-dominated regime. This opens a wide range of options for optimizing the extraction-recombination balance. We also show that energetic disorder affects charge encounter more than CT dissociation. As a consequence, decreasing the disorder increases $\mu$ stronger than $k_2$. On the other hand, for a given mobility, a larger disorder and hence a larger hopping rate is preferred when aiming at highly suppressed recombination. Our simulations suggest energetic disorder as an important parameter to be considered when aiming at a detailed understanding of non-geminate recombination in solar cells based on disordered semiconductors.

AUTHOR INFORMATION

Affiliations

**Institute for Physics and Astronomy, University of Potsdam, Germany**

Guangzheng Zuo, Safa Shoaee, Dieter Neher

**Centre for Advanced Materials, Ruprecht-Karls-Universität Heidelberg, Germany**

Martijn Kemerink

Corresponding author

Correspondence to Dieter Neher, Martijn Kemerink

Email: neher@uni-potsdam.de, martijn.kemerink@cam.uni-heidelberg.de

ACKNOWLEDGMENT

G.Z. thanks the Alexander von Humboldt Foundation for funding. S. S. and D.N. acknowledge funding from the Deutsche Forschungsgemeinschaft (DFG, German Research Foundation through the project Fabulous (NE 410/20, SH 1669/1-1). M. K. thanks the Carl Zeiss Foundation for financial support. We also thank Sebastian Wilken (Åbo Akademi University) as well as Oskar Sandberg and Ardalan Armin (Swansea University) for their careful reading of the manuscript and for fruitful discussion.

REFERENCES

[1] C. Li, J. Zhou, J. Song, J. Xu, H. Zhang, X. Zhang, J. Guo, L. Zhu, D. Wei, G. Han, J. Min, Y. Zhang, Z. Xie, Y. Yi, H. Yan, F. Gao, F. Liu, and Y. Sun, *Non-Fullerene Acceptors with Branched Side Chains and Improved Molecular Packing to Exceed 18% Efficiency in Organic Solar Cells*, Nat. Energy **6**, 605 (2021).

[2] Q. Liu, Y. Jiang, K. Jin, J. Qin, J. Xu, W. Li, J. Xiong, J. Liu, Z. Xiao, K. Sun, S. Yang, X. Zhang, and L. Ding, *18% Efficiency Organic Solar Cells*, Sci. Bull. **65**, 272 (2020).

[3] F. Liu, L. Zhou, W. Liu, Z. Zhou, Q. Yue, W. Zheng, R. Sun, W. Liu, S. Xu, H. Fan, L. Feng, Y. Yi, W. Zhang, and X. Zhu, *Organic Solar Cells with 18% Efficiency Enabled by an Alloy Acceptor: A Two-in-One Strategy*, Adv. Mater. **33**, 2100830 (2021).

[4] Y. Cui, H. Yao, J. Zhang, K. Xian, T. Zhang, L. Hong, Y. Wang, Y. Xu, K. Ma, C. An, C. He, Z. Wei, F. Gao, and J. Hou, *Single-Junction Organic Photovoltaic Cells with Approaching 18% Efficiency*, Adv. Mater. **32**, 1908205 (2020).

[5] M. Zhang, L. Zhu, G. Zhou, T. Hao, C. Qiu, Z. Zhao, Q. Hu, B. W. Larson, H. Zhu, Z. Ma, Z. Tang, W. Feng, Y. Zhang, T. P. Russell, and F. Liu, *Single-Layered Organic Photovoltaics with Double Cascading Charge Transport Pathways: 18% Efficiencies*, Nat. Commun. **12**, 1 (2021).

[6] A. Karki, A. J. Gillett, R. H. Friend, and T. Q. Nguyen, *The Path to 20% Power Conversion Efficiencies in Nonfullerene Acceptor Organic Solar Cells*, Adv. Energy Mater. **11**, 1 (2021).

[7] D. Neher, J. Kniepert, A. Elimelech, and L. J. A. Koster, *A New Figure of Merit for Organic Solar Cells with Transport-Limited Photocurrents*, Sci. Rep. **6**, 24861 (2016).

[8] S. Wilken, D. Scheunemann, S. Dahlström, M. Nyman, J. Parisi, and R. Österbacka, *How to Reduce Charge Recombination in Organic Solar Cells: There Are Still Lessons to*
27

# General Rules for the Impact of Energetic Disorder and Mobility on Nongeminate Recombination in Phase-Separated Organic Solar Cells

*Guangzheng Zuo[1], Safa Shoaee[1], Martijn Kemerink[2], Dieter Neher[1]\**

[1]Dr. G. Zuo, Prof. S. Shoaee, Prof. D. Neher, Institute for Physics and Astronomy, University of Potsdam, 14476 Potsdam-Golm, Germany
E-mail: neher@uni-potsdam.de

[2]Prof. Dr. M. Kemerink, Centre for Advanced Materials (CAM), Ruprecht-Karls-Universität Heidelberg, 69120 Heidelberg, Germany
Email: martijn.kemerink@cam.uni-heidelberg.de

**Details of the kMC simulation**

Charge transport is described as a hopping process based on Miller–Abrahams rates, in which the hopping with rate $\nu_{ij}$ from an initial state *i* with energy $E_i$ to a final state *j* with energy $E_j$, separated by a distance $r_{ij}$ is given by

$$\nu_{ij} = \begin{cases} \nu_0'(r_{ij}) \exp\left(-\dfrac{\Delta E}{k_B T}\right), & \Delta E > 0 \\ \nu_0'(r_{ij}), & \Delta E \leq 0 \end{cases}$$

where $\nu_0'$ is the rate of downward hopping and $\Delta E = E_j - E_i \pm q\vec{r}_{ij} \cdot \vec{F} + \Delta E_C$, with $\vec{F}$ is the external electric field, $\vec{r}_{ij}$ the vector connecting initial and final sites, and $q$ the positive elementary charge. The + (−) sign refers to electron (hole) hopping. The term $\Delta E_C$ is the change in Coulomb energy, calculated by explicit evaluation of the interaction of the moving charge with (a) all other charges in the simulated device and (b) their image charges, as well as of the interaction of the image charges of the moving particle with (c) the particle itself and (d) all other particles. In order to avoid divergences at zero separation, the Coulomb interaction between a pair of (unlike) charges with charges $E_C = -q/4\pi\varepsilon_0\varepsilon_r r_{eh}$ with $\varepsilon_0\varepsilon_r$ the dielectric constant ($\varepsilon_r$ = 3.6) and $r_{eh}$ the electron-hole distance is truncated at minus the approximate exciton binding energy for which we choose $E_b^{ex}$ = 0.5 eV. $k_B T$ is the thermal energy, the prefactor $\nu_0'$ is corrected for the differences in tunnelling distance $r_{ij}$ as

$$\nu_0' = \nu_0 \exp\left(-2\alpha(r_{ij} - a_{NN})\right)$$

with $\alpha$ the inverse localization radius.

The single-particle site energies $E_i$ are drawn from a Gaussian distribution function

$$f(E_i) = \frac{1}{\sqrt{2\pi\sigma^2}} \exp\left(-\frac{(E_i - E_0)^2}{2\sigma^2}\right)$$

with $E_0$ the mean energy and $\sigma$ the broadening of the total density of states (DOS) $N_0$. The HOMO and LUMO energy of a single site are assumed to be uncorrelated. For each initial site $i$, the number of possible final sites $j$ is set to 26, reflecting a 3 × 3 × 3 shell around the initial site.

Simulations are either performed in a homogeneous system, where each site can function as either donor or acceptor, or for a phase-separated morphology as described in the text. Free holes and electrons were considered as intial carriers, homogeneously distributed on the donor and acceptor sites. The positive and negative charge carriers can randomly form excitons or CT states before eventually leaving the system by recombination. Exciton diffusion by the Förster resonant energy transfer (FRET) mechanism is explicitly accounted for. The transition rate is evaluated as

$$v_{ij}^F = v_{ex} \left(\frac{R_0}{r_{ij}}\right)^6$$

where $R_0$ is the Förster radius and $k_{ex}$ the radiative exciton decay rate. CT recombination with a rate $k_{CT}$ is allowed whenever a hole and an electron sit on neighboring sites. Since we are interested in intrinsic material properties, we used periodic boundary conditions in all directions (no contacts).

For recombination studies, the electric field is set to zero ($\vec{F} = 0$) equivalent to open circuit conditions. The transient value of the bimolecular recombination factor $k_2$ is extracted from the simulations by

$$R = \frac{dn_{CS}(t)}{dt} = k_2(t) \cdot n_{CS}^2(t)$$

where $n_{CS} = n_{total} - (n_{CT} + n_{ex})$. Here $n_{CS} = n_e = n_h$ is the density of charge separated states, $n_{total}$ is total density of electron-hole pairs and $n_{CT}$ and $n_{ex}$ is the densities of CT states and excitons, respectively.

For the mobility calculations, a finite but low electric field of 1x10$^7$ V/m was used, and all other parameters are the same as what in corresponding recombination calculations.

**Key parameters used in kMC simulation.**

| Parameter | Value |
| --- | --- |
| Box size [sites] | 30 x 30 x 56 |
| Inter site distance, $a_{NN}$ [nm] | 1.8 |
| Energetic disorder, $\sigma_{DOS,e} = \sigma_{DOS,h}$ [meV] | 65, 90, 120 |
| Attempt-to-hop frequency, $\nu_{0,e} = \nu_{0,h}$ [s$^{-1}$] | 1 x 10$^8$ to 10$^{12}$ |
| Rate of CT state recombination, $k_{CT}$ [s$^{-1}$] | 1 x 10$^8$ or 1 x 10$^7$ |
| Rate of exciton recombination, $k_{ex}$ [s$^{-1}$] | 1 x 10$^9$ |
| Inverse localization length, $\alpha$ [m$^{-1}$] | 1.67 x 10$^9$ |
| Temperature, $T$ [K] | 300 |
| LUMO of donor, $E_{donor}^{LUMO}$ [eV] | 3.5 |
| HOMO of donor, $E_{donor}^{HOMO}$ [eV] | 5.5 |
| LUMO of acceptor, $E_{acceptor}^{LUMO}$ [eV] | 4.1 |
| HOMO of acceptor, $E_{acceptor}^{HOMO}$ [eV] | 5.9 |

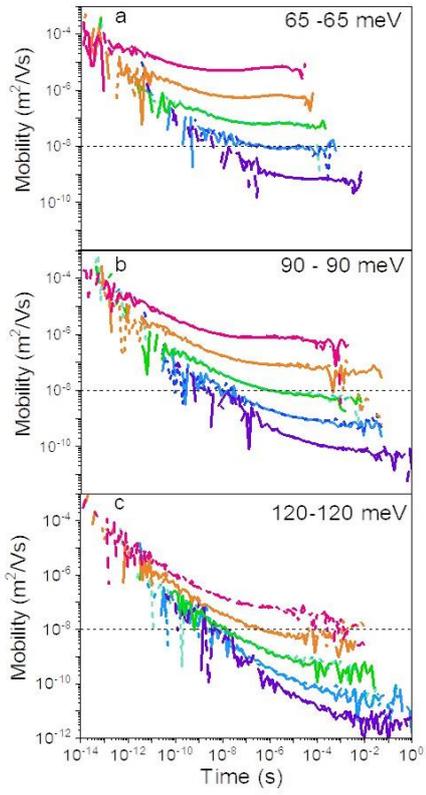
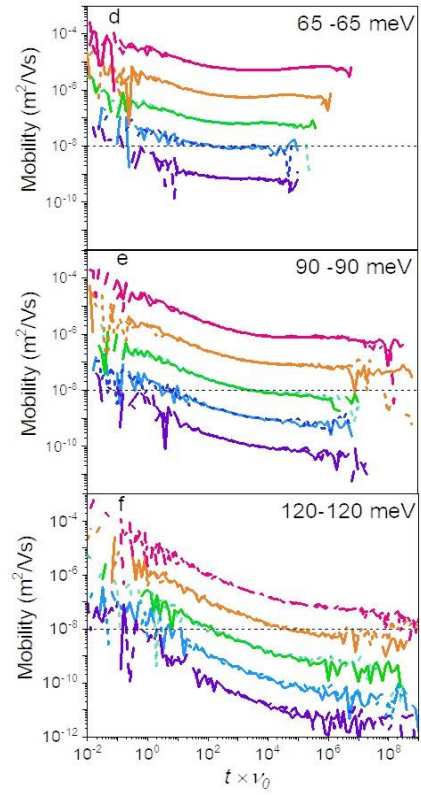
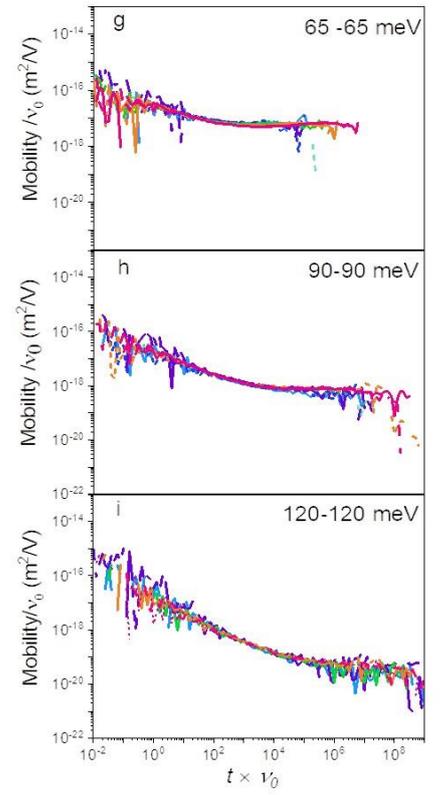
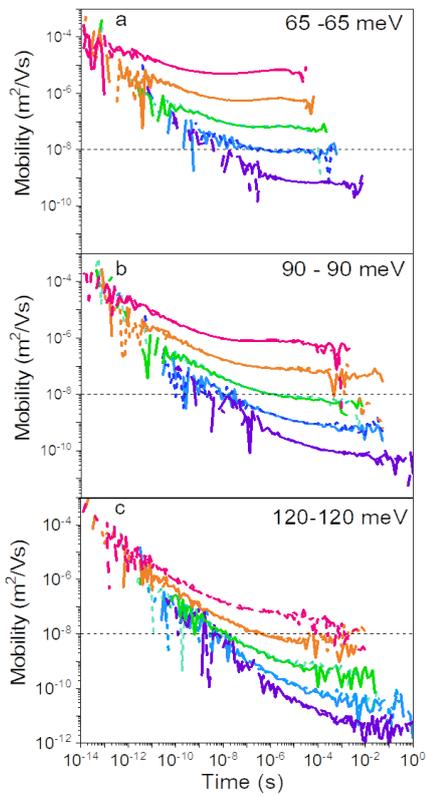
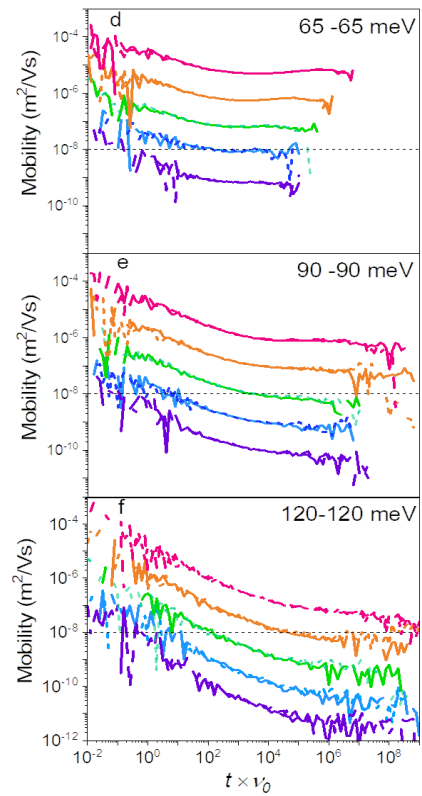
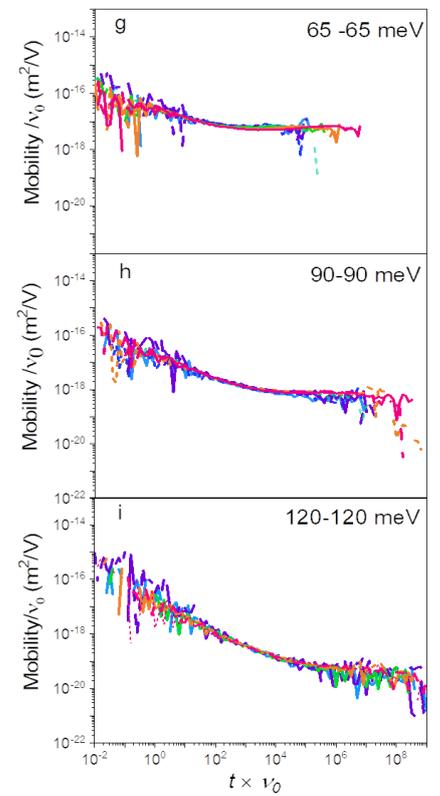

***Figure S1:** Transient mobility as function of time (**a, b, c**) and reduced time ($t \times \nu_0$) (**d, e, f**) for a phase-separated blend for variable energetic disorder and hopping frequencies under a voltage of 1V. Universal mobility plot (**g, h, i**) with both mobility ($\mu/\nu_0$) and time ($t \times \nu_0$) modified by attempt to hop frequency $\nu_0$. Simulations were performed for $\sigma_{DOS,e} = \sigma_{DOS,h} =$ 65 (a, d, g), 90 (b, e, h) and 120 meV (c, f, i), T = 300K, $a_{NN}$ = 1.8nm, $k_{CT}$ ($k_f$) = $1\times10^8$ $s^{-1}$, $\nu_{0,e} = \nu_{0,h} = 1\times10^8, 10^9, 10^{10}, 10^{11}$ and $10^{12}$ $s^{-1}$ (purple to red), respectively.*

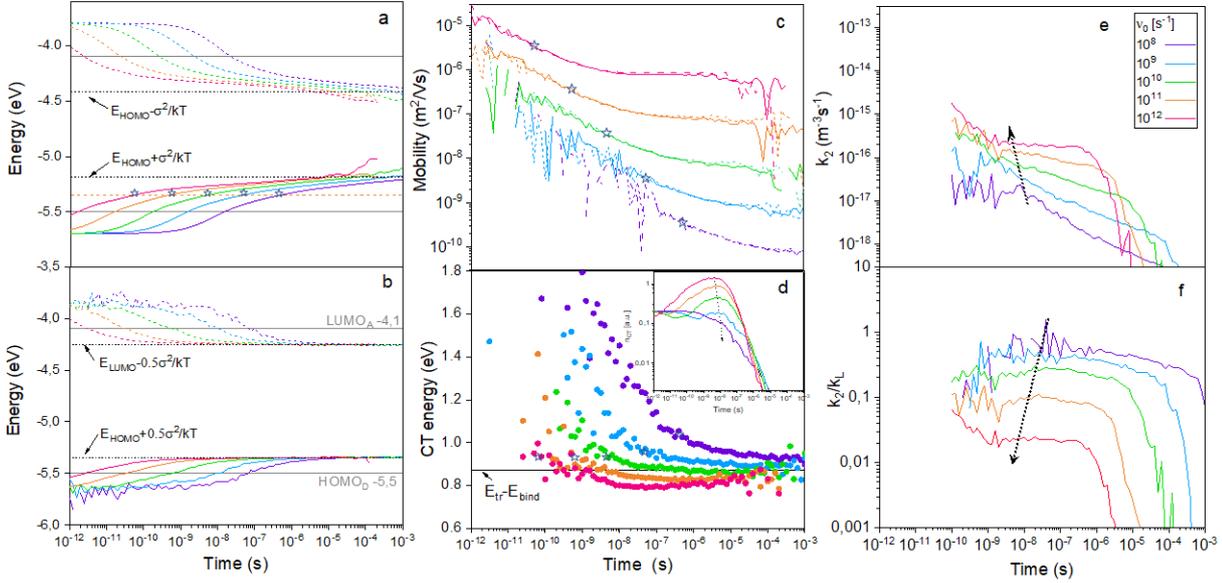

**Figure S2:** *Simulated transients for an energetic disorder $\sigma_{DOS,e} = \sigma_{DOS,h} = 90\ meV$ of (a) the mean energy the charges, (b) the energy of moving charges, (c) the bipolar mobilities for holes (solid lines) and electrons (dashed lines), (d) the energy of recombining CT states (e) of $k_2$ and (f) of the reduction factor for variable attempt-to-hop frequencies plotted as function of time (see Figure 3 for the corresponding plot as function of $t \times \nu_o$. The kMC parameters are the same as in Figure 3a. The star symbol marks the time when the mean energy crosses the effective transport energy $E_{tr} = E \mp \sigma_{DOS}^2/2k_BT$ (the grey horizontal dashed guideline) for different $\nu_o$, and the red horizontal guideline shows the equilibrium energy $E_\infty = E \mp \sigma_{DOS}^2/k_BT$, where (−) for electrons and (+) for holes. $E_{G,tr} - E_{bind} = \left(E_{LUMO} - \frac{\sigma_{DOS,e}^2}{2k_BT}\right) - \left(E_{HOMO} + \frac{\sigma_{DOS,h}^2}{2k_BT}\right) - E_{bind}$. Arrow indicates the maximum of the CT population.*

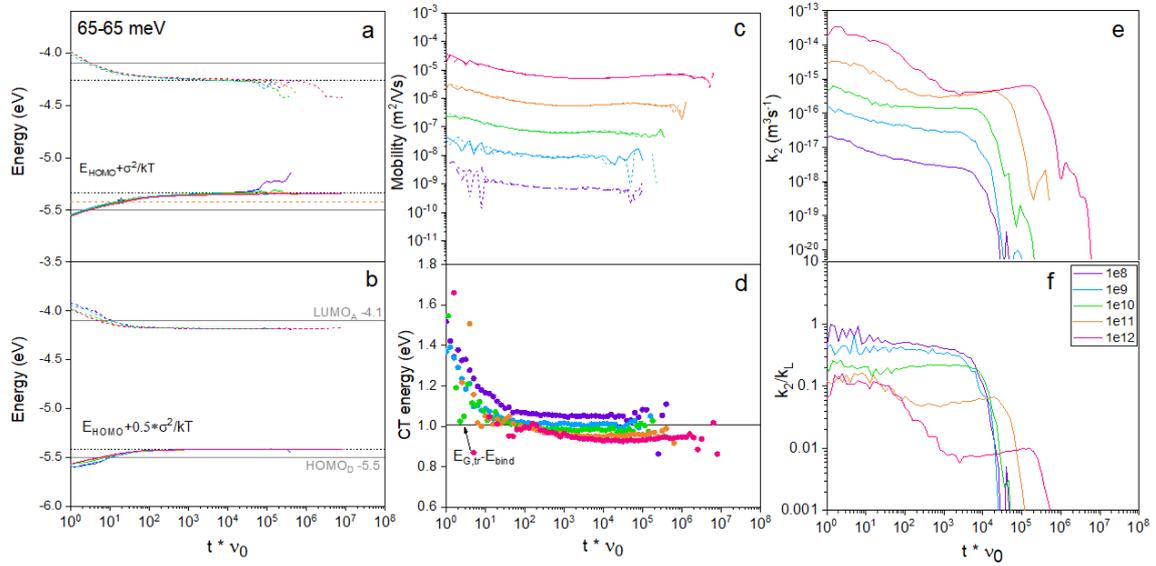

*Figure S3*: Simulated transients for an energetic disorder $\sigma_{DOS,e} = \sigma_{DOS,h} = 65\ meV$ of (a) the mean energy the charges, (b) the energy of moving charges, (c) the bipolar mobilities for holes (solid lines) and electrons (dashed lines), (d) the energy of recombining CT states (e) of $k_2$ and (f) of the reduction factor for variable attempt-to-hop frequencies plotted as function of $t \times \nu_o$. The kMC parameters as well as the meaning of all symbols and lines are the same as in Figure 3 expect the energetic disorder.

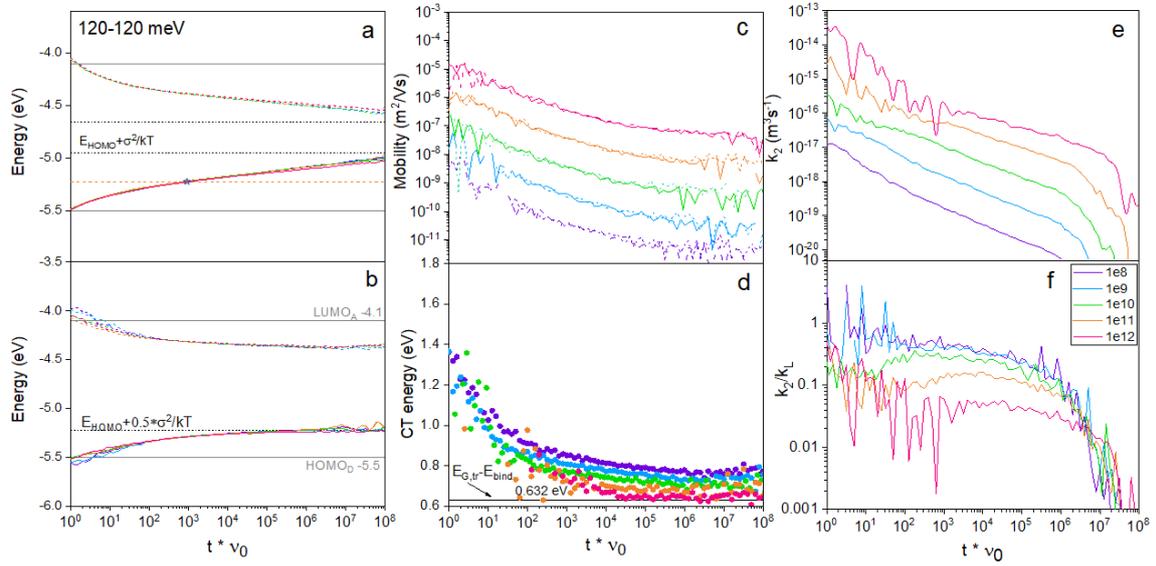

*Figure S4*: Simulated transients for an energetic disorder $\sigma_{DOS,e} = \sigma_{DOS,h}$ = 120 meV of (a) the mean energy the charges, (b) the energy of moving charges, (c) the bipolar mobilities for holes (solid lines) and electrons (dashed lines), (d) the energy of recombining CT states (e) of $k_2$ and (f) of the reduction factor for variable attempt-to-hop frequencies plotted as function of $t \times \nu_o$. The kMC parameters as well as the meaning of all symbols and lines are the same as in Figure 3 expect the energetic disorder.

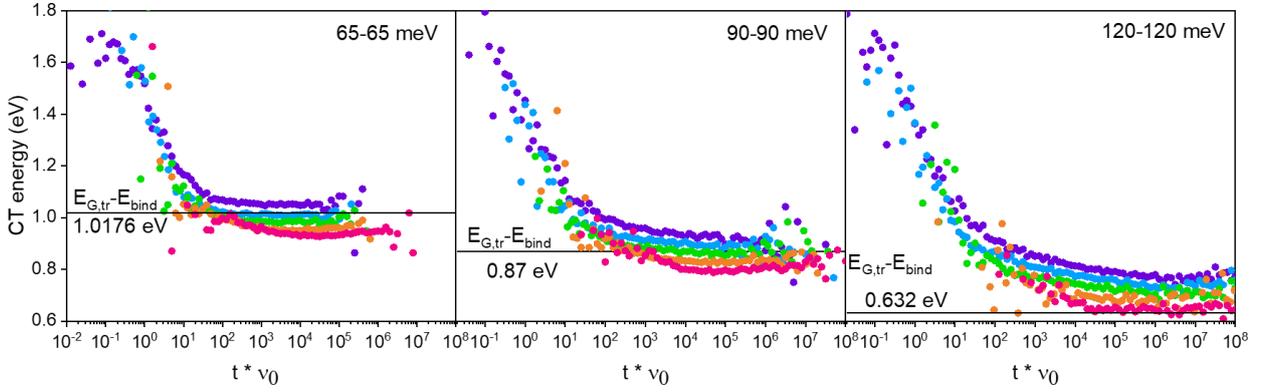

*Figure S5: Dynamic of the energy of recombining CT states as function of the reduced time for variable energetic disorder and hopping frequencies.* $E_{G,tr} - E_{bind} = \left(E_{LUMO} - \frac{\sigma^2_{DOS,e}}{2k_BT}\right) - \left(E_{HOMO} + \frac{\sigma^2_{DOS,h}}{2k_BT}\right) - E_{bind}$. *Expect the energetic disorder, the kMC parameters are the same as in Figure 2.*

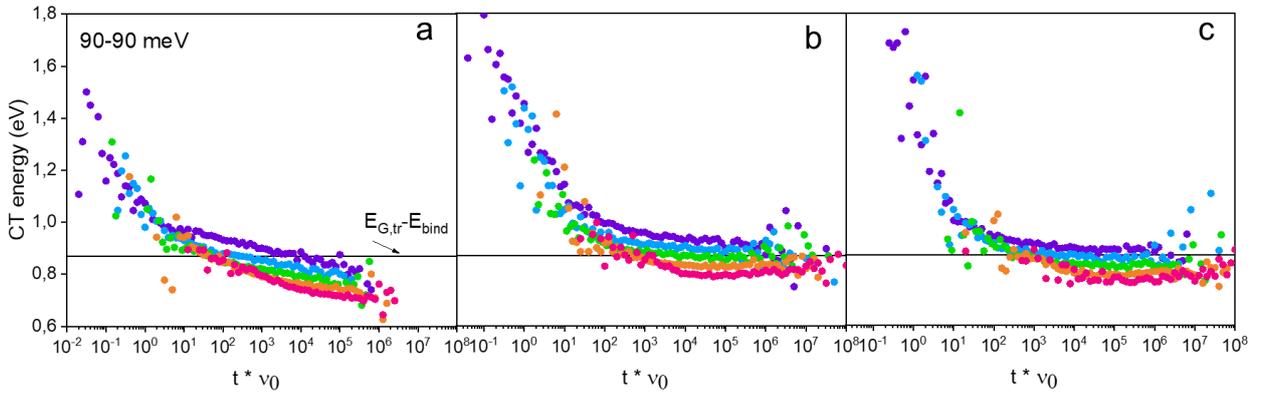

*Figure S6: Dynamic of the energy of recombining CT states as function of the reduced time for variable energetic disorder and hopping frequencies* for a homogenous blend **(a)** and a phase separated blend with $k_f = 1 \times 10^8 \ s^{-1}$ **(b)** and $k_f = 1 \times 10^7 \ s^{-1}$ **(c)**. $E_{G,tr} - E_{bind} = \left(E_{LUMO} - \frac{\sigma^2_{DOS,e}}{2k_BT}\right) - \left(E_{HOMO} + \frac{\sigma^2_{DOS,h}}{2k_BT}\right) - E_{bind}$.

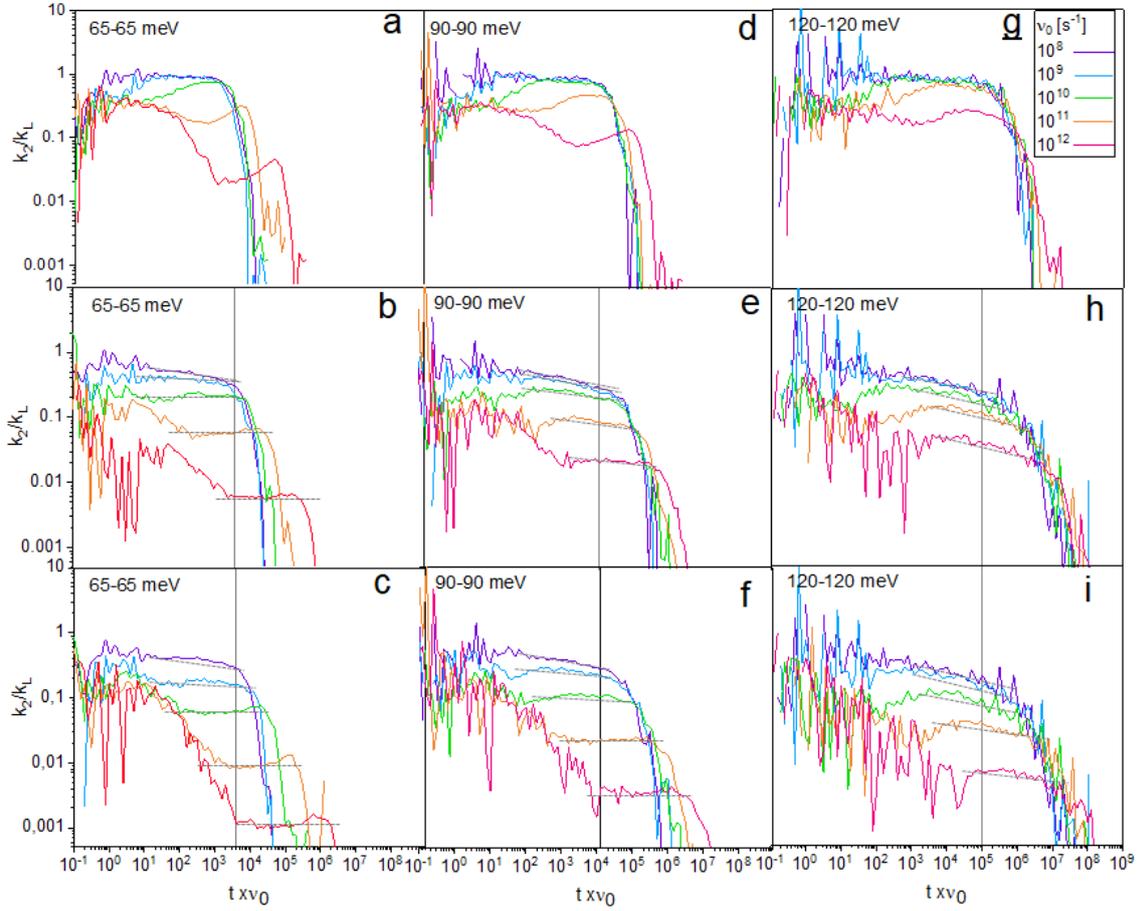

*Figure S7:* Reduction factor $\gamma = k_2/k_L$ as function of the reduced time for either a homogeneous blend with $k_f = 1 \times 10^8\ s^{-1}$ *(a, d, g)*, or a phase-separated blend with $k_f = 1 \times 10^8\ s^{-1}$ *(b, e, h)* or $k_f = 1 \times 10^7\ s^{-1}$ *(c, f, i)*. The grey dashed lines highlight the regions of (nearly) constant $\gamma$, from which the data in **Figure 5** and **Figure 6** were deduced at the times marked by the solid vertical lines.

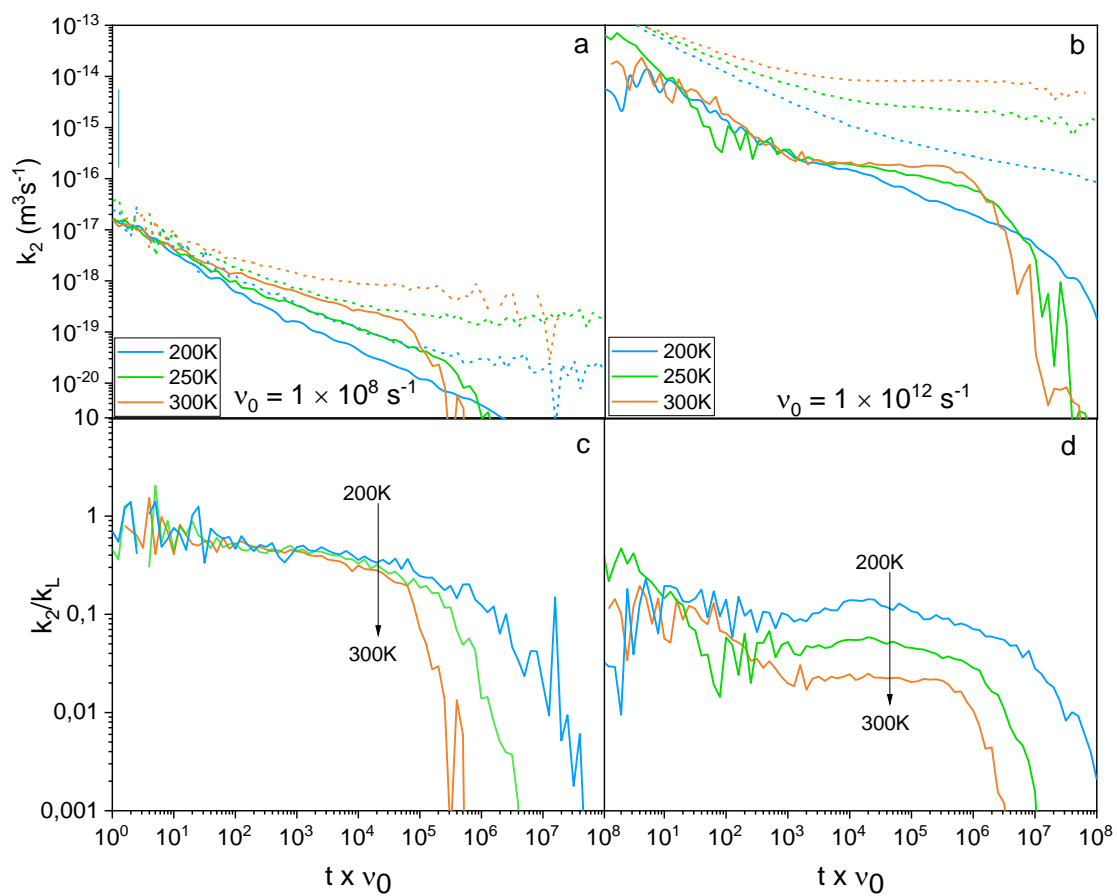

***Figure S8:*** *Simulated transients of $k_L$ (dashed lines) and $k_2$ (solid lines) **(a,b)** and the reduction factor $\gamma$ **(c,d)** as function of the reduced time for $\nu_0 = 10^8\ s^{-1}$ **(a, c)** and $\nu_0 = 10^{12}\ s^{-1}$ **(b,d)** and different temperatures. $\sigma_{DOS} = 90\ meV$ and $k_f = 10^8\ s^{-1}$ in all cases.*

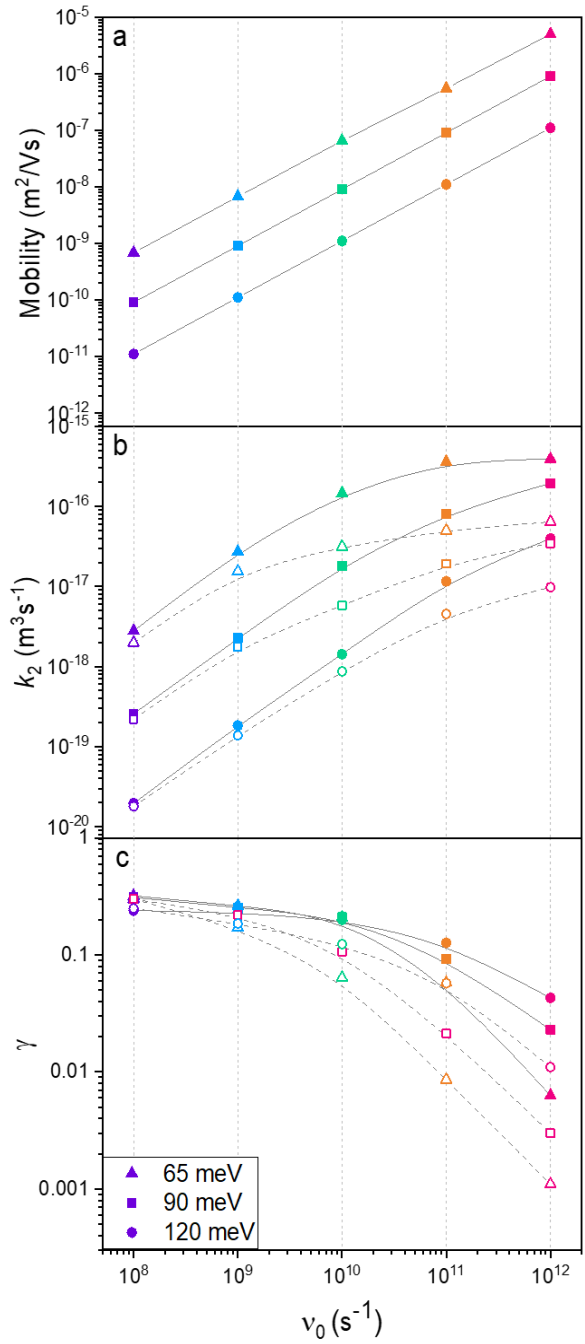

***Figure S9:*** *Simulated values of mobility (**a**), k₂ (**b**) and the reduction factor γ (**c**) for (nearly) equilibrated carriers, plotted as function of hopping frequency for different energetic disorder, $\sigma_{DOS} = 65\ meV, 90\ meV$ and $120\ meV$ with $k_f = 10^8\ s^{-1}$ (solid lines) and $k_f = 10^7\ s^{-1}$ (dashed lines), respectively.*